\documentclass{article}
\usepackage{spconf,amsmath,graphicx,makecell,boldline,subfigure,float,algpseudocode,algorithm,amsfonts}
\usepackage[pagebackref=true,breaklinks=true,colorlinks,bookmarks=false]{hyperref}

\title{SCORE-BASED DIFFUSION MODELS FOR PHOTOACOUSTIC TOMOGRAPHY IMAGE RECONSTRUCTION}
\makeatletter
\def\@name{ \emph{Sreemanti Dey, Snigdha Saha, Berthy T. Feng, Manxiu Cui, Laure Delisle,} \\ \emph{Oscar Leong, Lihong V. Wang, Katherine L. Bouman}}
\makeatother
\address{California Institute of Technology}
\begin{document}
%
\maketitle

\begin{abstract}
Photoacoustic tomography (PAT) is a rapidly-evolving medical imaging modality that combines optical absorption contrast with ultrasound imaging depth.  
One challenge in PAT is image reconstruction with inadequate acoustic signals due to limited sensor coverage or due to the density of the transducer array. Such cases call for solving an ill-posed inverse reconstruction problem. 
In this work, we use score-based diffusion models to solve the inverse problem of reconstructing an image from limited PAT measurements. The proposed approach allows us to incorporate an expressive prior learned by a diffusion model on simulated vessel structures while still being robust to varying transducer sparsity conditions.  
\end{abstract}
\section{INTRODUCTION}
\vspace{-0.1in}
\label{sec:intro}
Photoacoustic tomography (PAT) is a low-cost, ionizing-radiation-free technique for medical imaging. As such, it is growing in popularity and used in practical applications such as diagnosing breast cancer \cite{lin2018single}. PAT measurements are sensor signals from a transducer array surrounding the object of interest, which then must be reconstructed into a human-interpretable image. However, physical and resource limitations may make it impossible to fully encompass the object with transducers (limited-view problem \cite{limited_view}) or build a dense-enough array to prevent aliasing (spatial-aliasing problem \cite{spatial_aliasing1,spatial_aliasing2}), limiting the reliability of a direct inversion. 

With inadequate measurements, PAT image reconstruction can be formulated as an ill-posed inverse problem. Backprojection is a traditional solution but incorporates no priors and is prone to artifact-heavy reconstructions \cite{rosenthal2011immi}. Model-based methods combine the measurement forward model with an image regularizer, but they
do not capture complex image statistics, resulting in unrealistic reconstructions \cite{liu2016curvedriven}.

Deep learning poses an opportunity to incorporate more sophisticated image priors into the reconstruction. However, current deep-learning approaches are supervised with paired training data \cite{davoudi2019sau} and thus do not generalize to all measurement conditions. Practical applications call for a deep-learning approach that can be flexibly used in different settings.

Diffusion models are state-of-the-art generative models that have achieved success on various inverse imaging problems \cite{kawar2022denoising,abuhussein2022adir,wang2022zero,chung2023diffusion,feng2023score}. Song et al.~\cite{song2022solving} introduced a way to condition the generated images of a trained diffusion model on compressed-sensing measurements obtained for MRI or CT. However, PAT image reconstruction is not a compressed-sensing problem and instead involves dense, highly-correlated, time-varying measurements.

We introduce a method for PAT image reconstruction using a trained diffusion model. Our method is inspired by Song et al.~\cite{song2022solving} but generalizes to \textit{any} type of linear inverse problem. We validate our approach on synthetic vascular structure images under different measurement conditions, including quantitative and qualitative comparisons to a supervised deep-learning method \cite{davoudi2019sau} and total-variation (TV) regularization \cite{becktvfista}. Our work offers a technical contribution by proposing a new technique for solving general linear inverse problems with diffusion models, as well as a practical contribution by demonstrating the utility of diffusion models for PAT imaging.
\vspace{-0.3in}
\section{BACKGROUND}
\label{sec:background}
\subsection{Photoacoustic tomography (PAT)}
\begin{figure}[htb]
  \centering
  \includegraphics[width=0.3\textwidth, trim={2.8cm 5.5cm 17.5cm 5cm}]{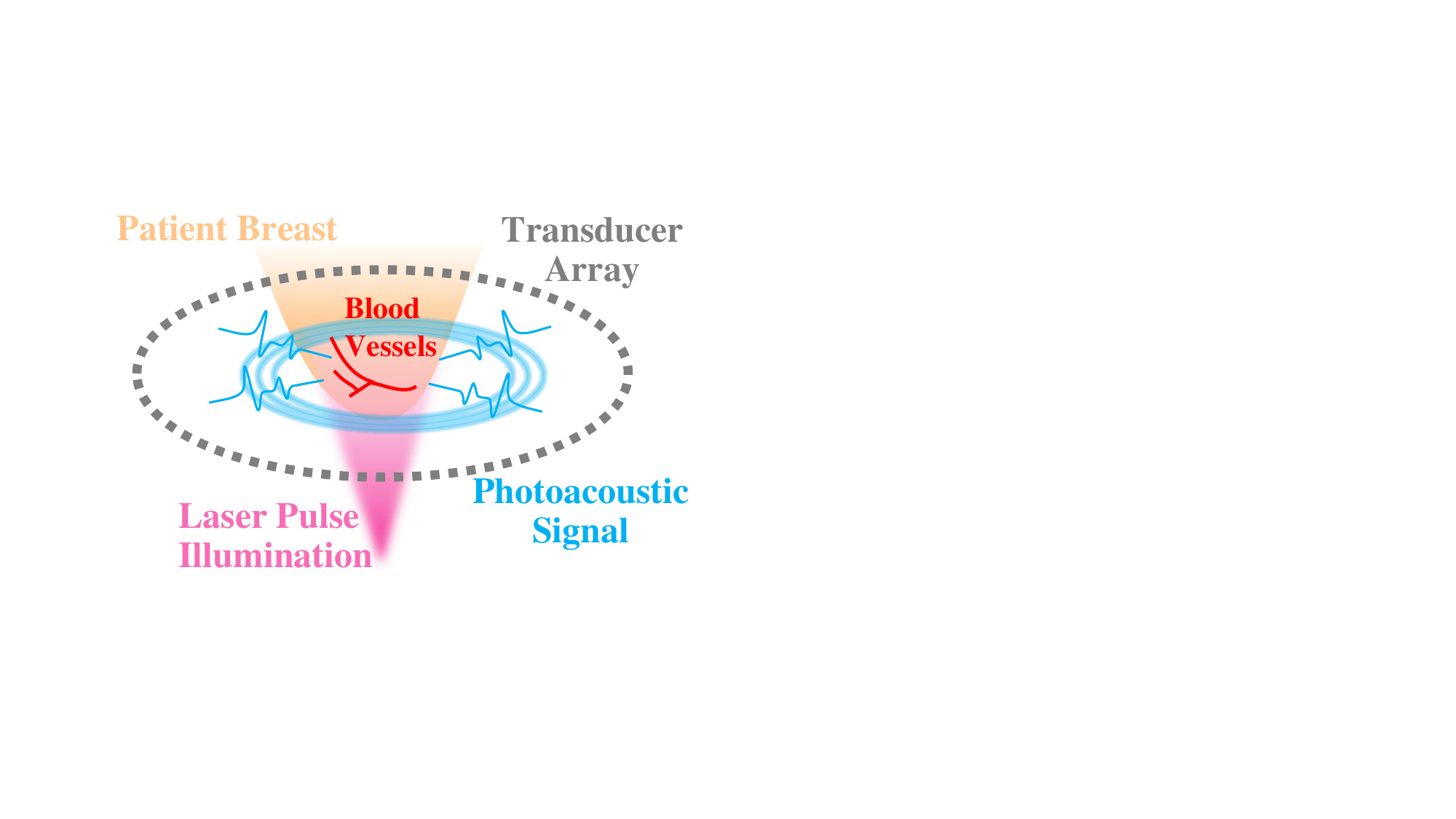}
\caption{\small{PAT measurement acquisition. A ring of ultrasound sensors (transducer array) surrounds the object to be imaged. 
The transducer array receives photoacoustic signals emitted in response to a laser pulse.}}
\label{fig:pat}
\end{figure}

PAT imaging maps optical absorption in scattering tissues with only surface-level measurements. In this work, we consider a ring-array-based PAT system for imaging a human breast. The breast of the patient is placed inside a ring of ultrasound sensors (Fig.~\ref{fig:pat}). Short-pulsed laser light incident on the patient's skin diffuses deep into the breast tissue, and then the absorbed energy by blood vessels generates ultrasonic waves as a result of thermoelastic expansion. The ultrasonic waves are detected by the transducers. This spatio-temporal data is used to reconstruct the image of the object \cite{STEINBERG201977}.

There are currently three major classes of image reconstruction methods for PAT: back-projection, time-reversal, and model-based. Back-projection methods do analytical inversion \cite{universal_backprojection} and, while fast and tractable, often lead to images with artifacts \cite{rosenthal2011immi}. Time-reversal methods, which use numerical simulations, give high-quality images but are computationally intensive \cite{razansky2012timerev}. Model-based methods minimize the difference between measured signals and predicted signals from an established forward model, often a linear operator \cite{liu2016curvedriven}. Model-based methods are becoming more common due to their independence from measurement geometry and balance between computation and quality. Our work similarly uses a linear forward model based on curve-driven model matrix inversion (CDMMI) \cite{liu2016curvedriven}. And while other deep-learning model-based methods exist (e.g., plug-and-play \cite{venkatakrishnan2013plug} and deep unrolling \cite{monga2021algorithm}), we leverage the strong prior of a diffusion model to achieve greater image quality. We note that concurrent work \cite{song2023photoacoustic} applies diffusion models to PAT with a focus on the spatial aliasing problem.

\subsection{Diffusion models for image reconstruction}
Diffusion models are state-of-the-art generative models that learn to sample from an image distribution \cite{ho2020denoising,kingma2021variational,nichol2021improved}. Recent methods have shown how to solve ill-posed inverse problems with a trained diffusion model as the prior \cite{song2022solving,kawar2022denoising,abuhussein2022adir,chung2023diffusion,feng2023score}, with most differing in the way measurements are incorporated into the sampling process of the diffusion model. Some of these methods have been applied to medical-imaging tasks like magnetic resonance imaging (MRI) \cite{song2022solving,jalal2021robust,feng2023efficient}, computed tomography (CT) \cite{song2022solving}, and ultrasound \cite{stevens2023dehazing}, but not to PAT. Our work builds upon an approach that has a simple projection step to incorporate measurements but was previously limited to compressed-sensing forward matrices \cite{song2022solving}. By generalizing the projection step to any linear forward model, we are able to address PAT.
\subsection{Score-based diffusion models}
Diffusion models learn to sample from an image distribution through gradual denoising. Score-based diffusion models model the process of adding noise to an image as a stochastic differential equation (SDE) \cite{song2021scorebased}:
\begin{equation}
\label{eq:forward_sde}
\mathrm{d}\mathbf{x}_t = \mathbf{f}(\mathbf{x}_t, t) \mathrm{d}t + g(t) \mathrm{d}\mathbf{w}_t,\quad t\in[0,T]
\end{equation}
where $\mathbf{x}_t\in\mathbb{R}^d$ is the image; $\mathbf{f}(\mathbf{x}_t,t)$ is the drift coefficient; $g(t)$ is the diffusion coefficient; and $\mathrm{d}\mathbf{w}_t$ is infinitesimal white noise. This SDE gives rise to a time-dependent distribution $p_t(\mathbf{x}_t)$. Higher time $t$ indicates more noise in $\mathbf{x}_t$.
We specifically use the Variance-Preserving (VP) SDE \cite{song2021scorebased} with $T=1$, which ensures that $p_T\approx\mathcal{N}(\mathbf{0},\mathbf{I})$.

Sampling from the clean distribution $p_0$ is based on the following \textit{reverse} SDE:
\begin{equation}
\label{eq:reverse_sde}
\mathrm{d}\mathbf{x}_t = \left[\mathbf{f}(\mathbf{x}_t, t) - g^2(t)\nabla_{\mathbf{x}_t}\log p_t(\mathbf{x}_t)\right]\mathrm{d}t + g(t)\mathrm{d}\mathbf{w}_t.
\end{equation}
Although the gradient $\nabla_{\mathbf{x}_t}\log p_t(\mathbf{x}_t)$ is unknown for an arbitrary image distribution $p_0$, it can be approximated with a convolutional neural network (CNN) called a \textit{score model} $\mathbf{s}_\theta$: $\mathbf{s}_\theta(\mathbf{x},t)\approx\nabla_{\mathbf{x}}\log p_t(\mathbf{x})$. The score model essentially learns to nudge images to higher probability.

Sampling 
starts with a noise image $\mathbf{x_T}\sim\mathcal{N}(\mathbf{0},\mathbf{I})$ that is gradually denoised by solving the reverse SDE (Eq.~\ref{eq:reverse_sde}) with $\nabla_{\mathbf{x}}\log p_t(\mathbf{x})$ replaced by $\mathbf{s}_\theta(\mathbf{x},t)$. Any numerical SDE solver can be used.
We use a second-order solver via Predictor-Corrector sampling \cite{song2021scorebased}.
\vspace{-0.1in}
\section{METHOD}
\label{sec:method}

\begin{figure}[htb]
  \centering
  \includegraphics[width=0.37\textwidth, trim={1cm 1cm 15.5cm 1cm}]{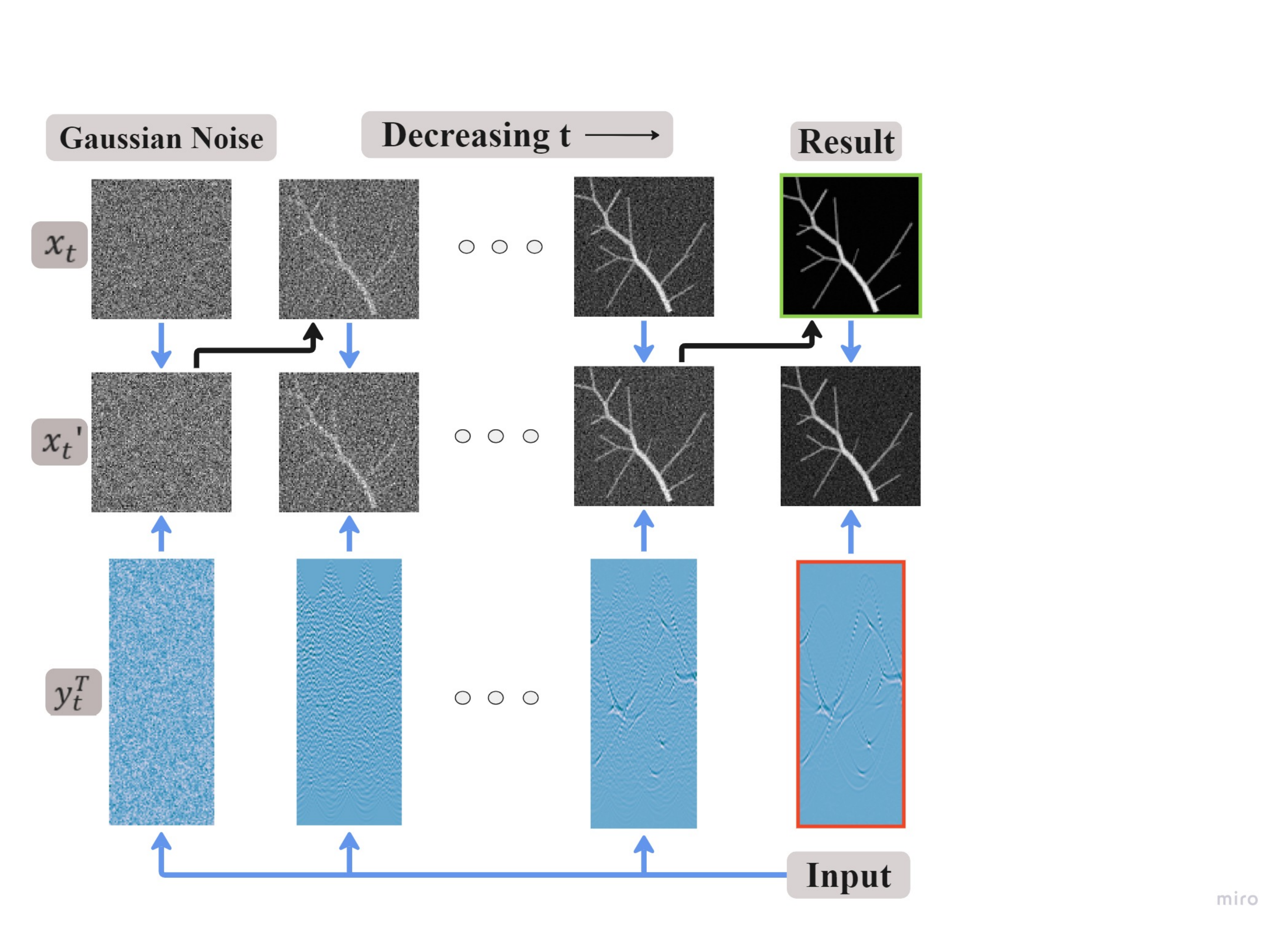}
\caption{\small{Our conditional sampling process with a trained diffusion model. Given PAT measurements, sampling starts with an image of Gaussian noise, which is transformed over many steps into the reconstructed PAT image. Each step involves a measurement-conditioning update (blue arrow) followed by a denoising update (black arrow) that takes the image closer to the learned prior.}}
\label{fig:model}
\end{figure}

Our approach adapts the unconditional sampling procedure to be conditioned on PAT measurements $\mathbf{y}$. Following Song et al.~\cite{song2022solving}, we model a diffusion process on $\mathbf{y}$ and at each diffusion time $t$, modify the image $\mathbf{x}_t$ to be consistent with the perturbed measurements $\mathbf{y}_t$. Song et al.~define the following measurement-conditioning step, which assumes an invertible matrix $\mathbf{T}$ and a measurement-reduction operator $\mathcal{P}(\mathbf{\Lambda})$:
\small
\begin{align}
\label{eq:prev_conditioning}
\mathbf{x}'_{t} \gets \mathbf{T}^{-1}[\lambda \mathbf{\Lambda} \mathcal{P}(\Lambda)\mathbf{y}_{t} + (1-\lambda)\mathbf{\Lambda} \mathbf{T}\mathbf{x}_{t} + (\mathbf{I} - \mathbf{\Lambda})\mathbf{T}\mathbf{x}_{t}].
\end{align}

\normalsize

Essentially, $\mathbf{x}'_t$ balances the image $\mathbf{x}_t$ produced by the unconditional diffusion model and the measurements $\mathbf{y}_t$, with $\lambda\in[0,1]$ tuning the weight of the measurements.

Our inverse problem, however, does not involve an invertible $\mathbf{T}$ matrix or subsampling operator $\mathcal{P}(\mathbf{\Lambda})$. Instead, our CDMMI forward matrix \cite{liu2016curvedriven} is an ill-conditioned tall matrix that produces highly-correlated measurements, making Eq.~\ref{eq:prev_conditioning} unusable.
We formulate a new measurement-conditioning step by solving a regularized maximum-likelihood objective:

\begin{align}
\label{eq:objective}
\mathbf{x}'_t &= \arg \min_{\mathbf{z} \in \mathbb{R}^d} \left[(1-\lambda)\lVert\mathbf{z} - \mathbf{x}_t\rVert_2^2 + \lambda\lVert \mathbf{y}_t - \mathbf{Az}\rVert_2^2\right] \\
&=\left(\lambda \mathbf{A}^\top\mathbf{A} + (1-\lambda) \mathbf{I}\right)^{-1}\left((1-\lambda)\mathbf{x}_{t} + \lambda \mathbf{A}^\top\mathbf{y}_{t}\right),\label{eq:condition}
\end{align}
where $\mathbf{A}$ is any forward matrix (in our case, the CDMMI matrix). Alg.~\ref{alg:reverse} details our conditional sampling procedure, which is visualized in Fig.~\ref{fig:model}.

\begin{algorithm}
\caption{\small{Our conditional sampling process, where $p_{0t}(\mathbf{y}_t \mid \mathbf{y})$ comes from the diffusion SDE (Eq.~\ref{eq:forward_sde}), and $\Delta \mathbf{x}'_t$ is the SDE solver output at time $t$ given $\mathbf{x}'_t$.}}
\small{
\label{alg:reverse}
\begin{algorithmic}
\Require $N, T, \mathbf{y}, \mathbf{A}, \lambda$
\State $t \gets T$, $\Delta t \gets -\frac{T}{N}$, $\mathbf{x}_T\sim\mathcal{N}(\mathbf{0},\mathbf{I})$
\While{$t > 0$}
    \State $\mathbf{y}_t \sim p_{0t}(\mathbf{y}_t \mid\mathbf{y})$
    \State $\mathbf{x}'_{t} \gets \left(\lambda \mathbf{A}^\top\mathbf{A} + (1-\lambda) \mathbf{I}\right)^{-1}\left((1-\lambda)\mathbf{x}_{t} + \lambda \mathbf{A}^\top\mathbf{y}_{t}\right)$
    \State $\mathbf{x}_{t+\Delta t} \gets \mathbf{x}'_{t} + \Delta \mathbf{x}'_{t}$
    \State $t \gets t + \Delta t$
\EndWhile
\State \Return $\mathbf{x}_0$
\end{algorithmic}}
\end{algorithm}

\begin{figure}[htb]
\centering
\includegraphics[width=0.37\textwidth,trim={0.5cm 1cm 12cm 0}]{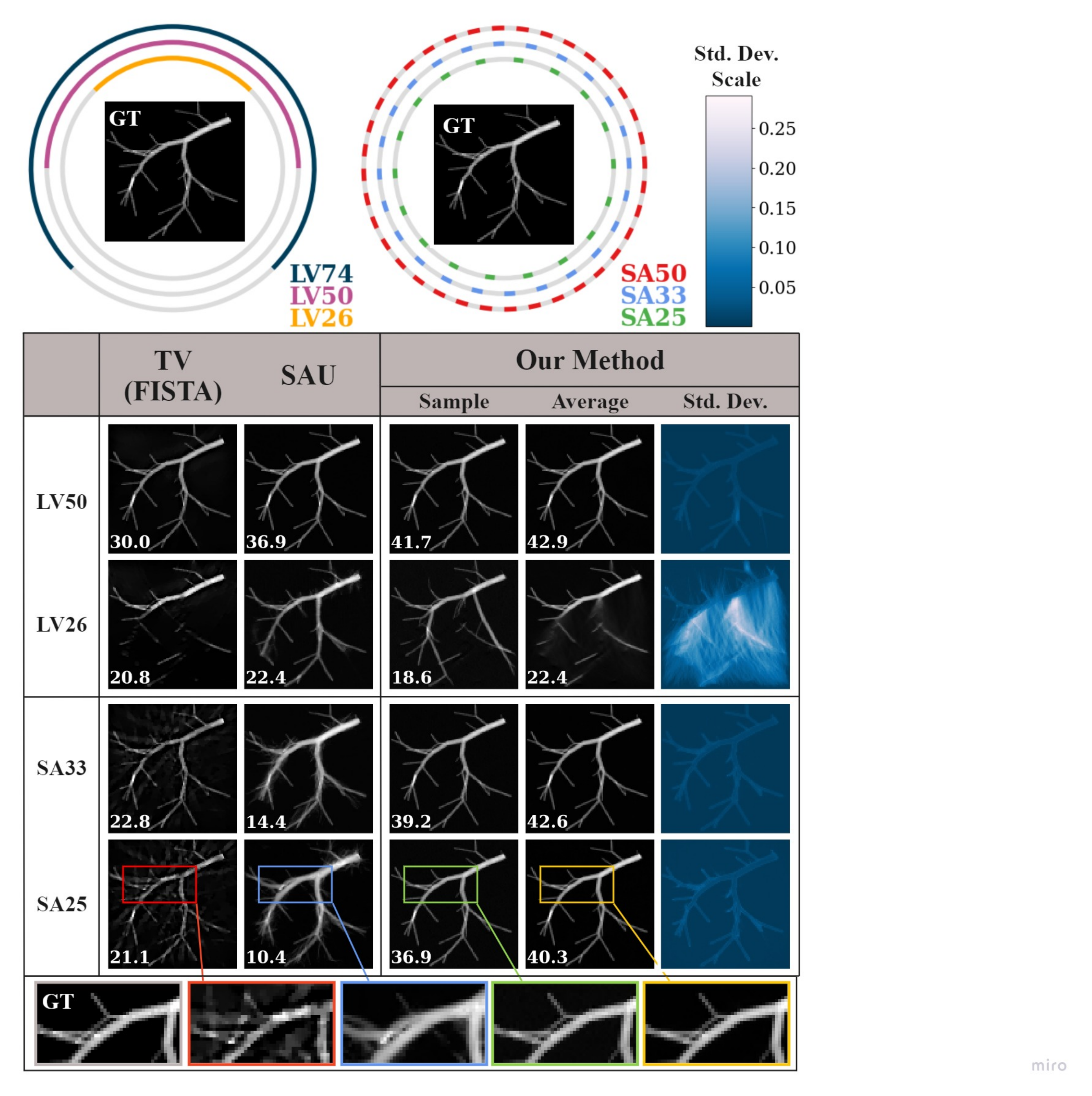}
\caption{
\small{Image reconstructions across meas.~settings. Top two diagrams illustrate the limited-view and spatial-aliasing configurations, resp. (e.g., ``LV74'' refers to limited-view with $74\%$ transducers). PSNR is on the bottom left of each image. Our method's results include one sample and the avg.~and std.~dev.~of 320 samples. The zoom-ins show high-fidelity details from our method that do not appear in baseline reconstructions. Overall, our method outperforms baselines in SA settings but may be prone to hallucination for LV (std.~dev.~maps show where hallucinations occur).
Qualitatively, our samples tend to appear closer to the prior. The mean of our samples generally outperforms baselines.}
}
\label{fig:reconstructions}
\end{figure}
\vspace{-0.3in}
\section{EXPERIMENTS}
\label{sec:experiments}
We performed experiments on simulated measurements of both synthetic vascular images and a real breast-tissue image. We compared to two baseline methods: (1) maximum-likelihood with total-variation regularization (\textbf{TV}) and (2) a fully-supervised deep-learning approach called Sparse Artefact U-Net (\textbf{SAU}) \cite{davoudi2019sau}. Optimization for (1) was done with the Fast Iterative Shrinkage-Thresholding Algorithm (FISTA) \cite{beck2009fista} \cite{becktvfista}. (2) uses a U-Net CNN \cite{ronneberger2015u} to learn the error between naive reconstructions and ground-truth images---it is important to note that with this baseline, separate models must be trained for each transducer configuration. To create the paired training data, we simulated measurements of the ground-truth images and used Tikhonov-regularized MLE for naive reconstructions from these measurements.

We created a dataset of synthetic vascular structure images with Vascusynth \cite{hamarneh2010vascusynth}, using the example parameters provided in the manual \cite{jassi2011vascusynth} but randomly setting the number of vascular nodes in each image.
9900 images were reserved for training our diffusion model and SAU, and 2000 images were reserved for validation of SAU.

We consider two types of limited sampling patterns of the transducer array: (1) limited view (LV)~\cite{limited_view} and (2) spatial aliasing (SA)~\cite{spatial_aliasing1}, illustrated in Fig.~\ref{fig:reconstructions}.
LV is more challenging, as only a portion of the image circumference is observable. SA  spaces the transducers equally around the circumference. We simulated measurements using a CDMMI forward matrix for each sampling configuration and added realistic Gaussian noise (30 dB SNR).

\subsection{Image-reconstruction quality}
\begin{figure}[htb]
  \centering
  \centerline{\includegraphics[width=0.5\textwidth]{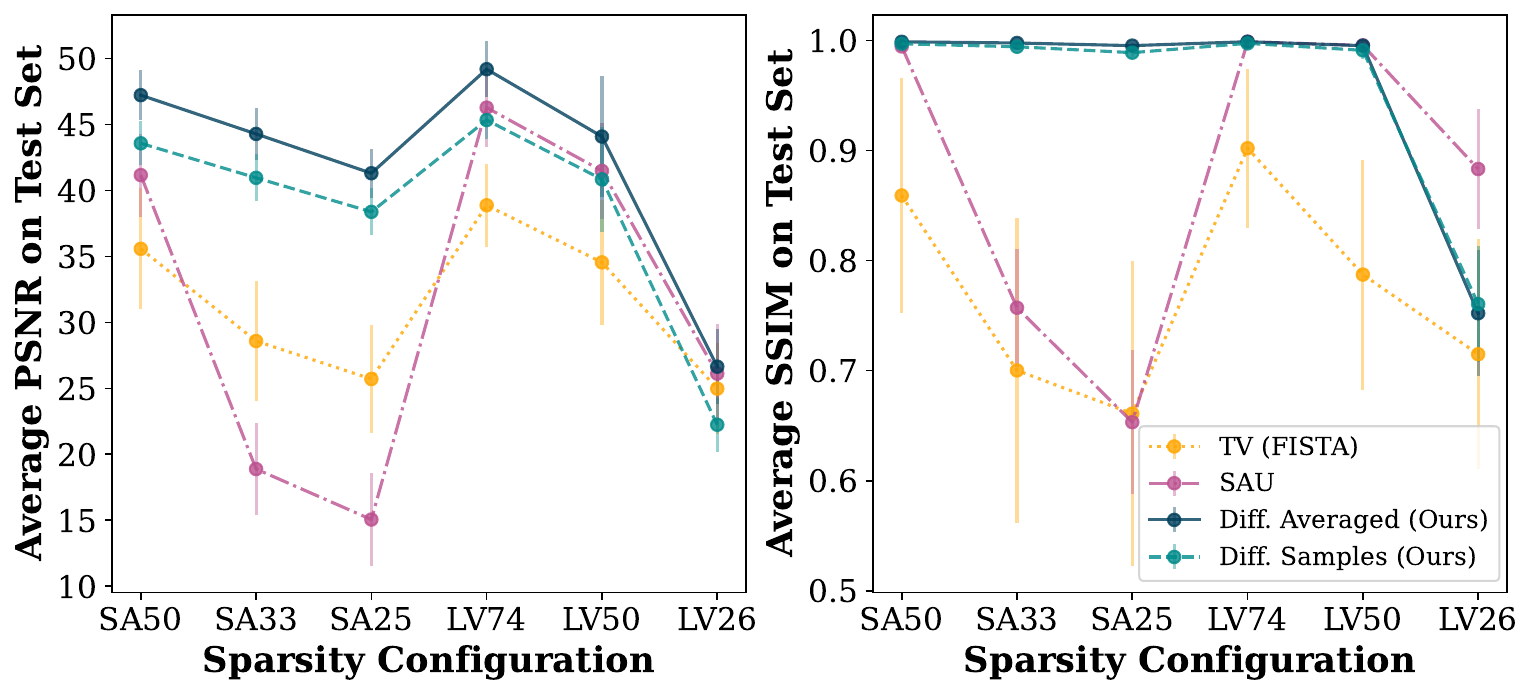}}
\caption{\small{Average PSNR and SSIM on the 10-image test set for TV, SAU, and our diffusion-model approach. 
``Diffusion Average" computes the avg.~PSNR or SSIM based on the empirical mean of 320 samples; ``Diffusion Sample" computes the avg.~PSNR or SSIM based on all samples for each measurement. 
Our samples beat both baselines for SA configurations and perform on par for LV, while our averaged reconstructions outperform baselines on nearly every configuration.}
}
\label{fig:psnrs}
\end{figure}

In Fig.~\ref{fig:psnrs}, we compare the average PSNR of reconstructions from our method versus those obtained with TV regularization and SAU. For SA, our method consistently outperforms the baselines with a notable PSNR improvement (e.g., $2.4$ to $23.3$ dB improvement over SAU across all sparsity levels). For LV, SAU achieves slightly higher PSNRs, but our approach still averages within one SAU standard deviation. Fig.~\ref{fig:reconstructions} shows reconstructions of an example test image. 

The diffusion model excels at generating images true to its learned prior, but this also means that it may hallucinate structures when given very limited measurements. In particular, we find that our PSNR performance is worse than SAU in the LV setting due to hallucination. Although qualitatively our image samples are more visually-convincing, certain features in the image should be cautiously interpreted. One way to assess the reliability of reconstructed features is to compute the empirical standard deviation of many samples from the conditional sampling process (Alg.~\ref{alg:reverse}). 

\subsection{Flexibility}
\begin{figure}[htb]
\centering
\includegraphics[width=0.37\textwidth,trim={1cm 1cm 25cm 1cm}]{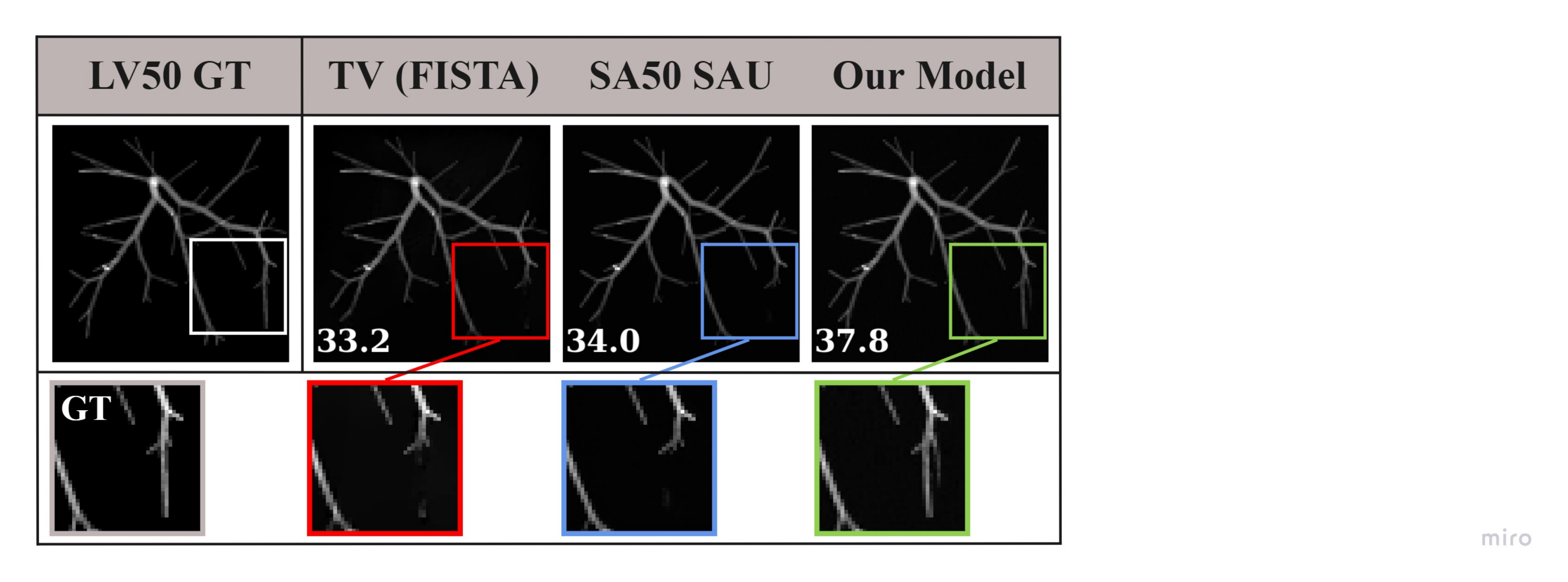}
\caption{\small{Comparison of SAU performance on a new transducer pattern vs.~our model and TV. Each model was tested on an LV50 image, but the SAU model used was trained on SA50 reconstructions. PSNRs show that SAU does not generalize across configurations. Zoom-in shows a GT feature that only our method was able to recover.}}
\label{fig:ood_measurements}
\end{figure}

We observe in Fig.~\ref{fig:ood_measurements} that transferring the same SAU to a different measurement setting results in significantly lower performance. 
In constrast, our method adapts to different settings without retraining.

\begin{figure}[htb]
    \centering
    \includegraphics[width=0.38\textwidth,trim={10cm 12.8cm 14cm 8cm},clip]{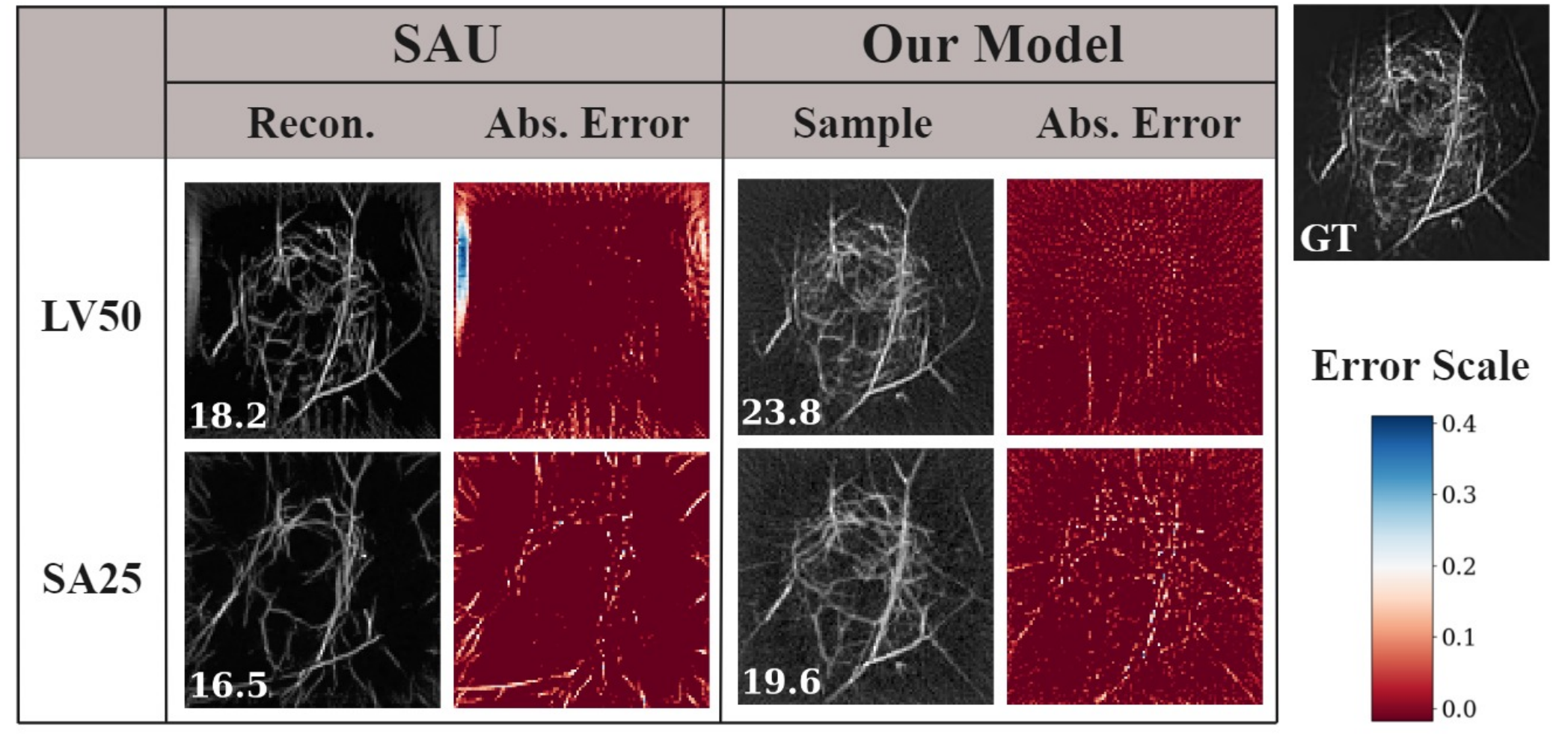}
    \caption{\small{Real breast image. Our diffusion model and SAU were each trained only on synthetic images. Ours produced higher-fidelity reconstructions for all configurations except LV26 (LV50 and SA25 shown here).}}
\label{fig:real}
\end{figure}

Our approach's flexibility also applies to out-of-distribution source images. Fig.~\ref{fig:real} shows that our method can plausibly reconstruct a breast image \cite{lin2018single} from simulated PAT measurements with good PSNR in all but the most extreme sparsity case, despite using a diffusion model trained only on synthetic vascular images.

\section{DISCUSSION}
\label{sec:discussion}
We have presented a method for unsupervised PAT image reconstruction using a trained diffusion model.
Our work builds upon a previous diffusion-model approach
both by proposing a new measurement-conditioning formula suitable for any linear forward model and by tackling the problem of PAT imaging. In our experiments with simulated measurements, we find that our method performs substantially better than traditional TV regularization and competitively to a fully-supervised deep-learning approach (and even better when taking the sample mean),
without requiring retraining for every transducer pattern. 
We also show better reconstruction of an out-of-distribution image of real breast tissue. Our work establishes a promising path to leveraging deep-learned priors for flexible photoacoustic tomographic imaging. 

\section{ACKNOWLEDGMENTS}
This work was supported in part by a Heritage Medical Research Fellowship Award, National Institutes of Health grants U01 EB029823 (BRAIN Initiative), and R35 CA220436 (Outstanding Investigator Award). L.W. has a financial interest in Microphotoacoustics, Inc., CalPACT, LLC, and Union Photoacoustic Technologies, Ltd., which, however, did not support this work. B.T.F. is supported by the NSF GRFP. M.C. would like to thank Yousuf Aborahama for fruitful discussions.

\footnotesize{
\bibliographystyle{IEEEbib}
\bibliography{refs.bib}
}

\end{document}